# Meron-Cluster Simulation of the $\theta$-Vacuum in the 2-d $O(3)$-Model *

W. Bietenholz$^a$, A. Pochinsky$^{a,b}$ and U.-J. Wiese$^a$

$^a$ Center for Theoretical Physics,
Laboratory for Nuclear Science, and Department of Physics
Massachusetts Institute of Technology
Cambridge, Massachusetts 02139, U.S.A.

$^b$ Institute for Theoretical and Experimental Physics
117259, Moscow B. Cheremushckinskaya 25, Russia

MIT Preprint, CTP 2433

May 24, 1995

## Abstract

The 2-d $O(3)$-model with a $\theta$-vacuum term is formulated in terms of Wolff clusters. Each cluster carries a half-integer topological charge. The clusters with charge $\pm 1/2$ are identified as merons. At $\theta = \pi$ the merons are bound in pairs inducing a second order phase transition at which the mass-gap vanishes. The construction of an improved estimator for the topological charge distribution makes numerical simulations of the phase transition feasible. The measured critical exponents agree with those of the $k = 1$ Wess-Zumino-Novikov-Witten model. Our results are consistent with Haldane's conjecture for 1-d antiferromagnetic quantum spin chains.

*This work is supported in part by funds provided by the U.S. Department of Energy (D.O.E.) under cooperative research agreement DE-FC02-94ER40818.

Some time ago Haldane conjectured [1] that integer and half-odd-integer 1-d antiferromagnetic quantum spin chains behave qualitatively differently. While integer spin chains have a mass-gap half-odd-integer chains should be gapless. This has been confirmed numerically for finite chains of spin 1 and 2 [2, 8] and analytically for half-odd-integer spins and for spin 1 [4]. The long-range physics of 1-d quantum spin chains is described by an effective 2-d classical $O(3)$-model. Haldane argued that the effective action for a chain of spins $S$ contains a topological term $i\theta Q$. Here $Q$ is the topological charge and $\theta = 2\pi S$ is the vacuum angle. Since the physics is periodic in $\theta$, i.e. $\theta \in ]-\pi, \pi]$, integer spins have $\theta = 0$ and half-odd-integer spins have $\theta = \pi$. The standard $O(3)$-model has a mass-gap in agreement with Haldane's conjecture. On the other hand, the conjecture implies that the mass-gap of the $O(3)$-model disappears at $\theta = \pi$. This corresponds to a phase transition in the vacuum angle. Due to the complex action it is notoriously difficult to simulate $\theta$-vacua numerically. A previous numerical study that was limited to $|\theta| < 0.8 \pi$ found no phase transition in that region [5]. In fact, Haldane's conjecture has not yet been verified in the context of the $O(3)$-model. In this paper we use the Wolff cluster algorithm [6] combined with a reweighting technique [7] to attack this problem. The construction of an improved estimator for the topological charge distribution enables us to simulate $\theta$-vacua reliably for any value of $\theta$.

Affleck and Haldane have suggested a dynamical mechanism that explains why the mass-gap disappears at $\theta = \pi$ [8]. In this picture pseudo-particles with topological charge $\pm 1/2$ — so-called merons — are the relevant degrees of freedom. At $\theta = 0$ the merons form an ideal gas. They disorder the system and thereby give non-zero mass to the physical particles. At $\theta = \pi$, on the other hand, the merons are bound in pairs and thus do not generate mass. Affleck confirmed this picture in a model where the $O(3)$ symmetry is explicitly broken to $O(2)$. Then the merons behave like vortices and the phase transition in $\theta$ is analogous to the Kosterlitz-Thouless transition of the $O(2)$-model. When the explicit $O(3)$ breaking is switched off it is unclear if this dynamical picture still holds. In fact there exists no definition of merons beyond the semiclassical approximation.

Here we formulate the $O(3)$-model in terms of Wolff clusters. Using an appropriate action on a triangular lattice each cluster has a uniquely defined half-integer topological charge. It turns out that most clusters are neutral, some have charges $\pm 1/2$ and very few carry larger charges. It is natural to identify merons as Wolff clusters with charge $\pm 1/2$. Indeed at $\theta = 0$ the Wolff clusters are completely independent, as they should in order to resemble merons. At $\theta = \pi$, on the other hand, no clusters with half-odd-integer charges persist. Instead they form bound pairs of integer charge. Hence the cluster formulation of the model provides a definition of merons beyond the semiclassical approximation, and it confirms Affleck's dynamical mechanism in the $O(3)$-model even without introducing symmetry breaking terms.

Actually Affleck's picture of the phase transition is more quantitative [9]. He argues that the model at $\theta = \pi$ is a conformal field theory in the same universality class as the $k = 1$ Wess-Zumino-Novikov-Witten model [10]. The critical exponents of this model are known analytically. The results of our numerical simulations are consistent with them, confirming Haldane's conjecture as well as Affleck's picture in the context of the $O(3)$-model.

For technical reasons we work on a 2-d triangular lattice with $V = 3L^2$ points. The lattice covers the volume $V$ of a regular hexagon with periodic boundary conditions. To each lattice site $x$ we attach a classical 3-component spin $\vec{e}_x \in S^2$ (of unit length). The action

$$S[\vec{e}] = \sum_{\langle xy \rangle} s(\vec{e}_x, \vec{e}_y) \, , \quad s(\vec{e}_x, \vec{e}_y) = \begin{cases} \frac{1}{g}(1 - \vec{e}_x \cdot \vec{e}_y) & \text{if } \vec{e}_x \cdot \vec{e}_y > -1/2 \\ \infty & \text{otherwise} \end{cases} \quad (1)$$

is a sum over nearest neighbor pairs $\langle xy \rangle$. It constrains neighboring spins to relative angles less than $2\pi/3$ which is necessary for our purposes as will be explained below. The spin field defines a mapping from the lattice to the sphere $S^2$. Using an appropriate interpolation, these mappings fall into topological classes characterized by an integer valued topological charge $Q \in \mathbf{Z}$. The charge $Q$ counts how many times $S^2$ is covered by the mapping. The total charge

$$Q[\vec{e}] = \sum_{\langle xyz \rangle} q(\vec{e}_x, \vec{e}_y, \vec{e}_z) \, , \quad q(\vec{e}_x, \vec{e}_y, \vec{e}_z) = A(\vec{e}_x, \vec{e}_y, \vec{e}_z)/4\pi, \quad (2)$$

is a sum of contributions from oriented elementary lattice triangles $\langle xyz \rangle$. The spins $\vec{e}_x$, $\vec{e}_y$ and $\vec{e}_z$ at the three corners $x$, $y$ and $z$ of a lattice triangle define a minimal spherical triangle on the sphere $S^2$. Each lattice triangle contributes the area $A(\vec{e}_x, \vec{e}_y, \vec{e}_z)$ of the corresponding spherical triangle divided by the area of $S^2$ with a sign depending on its orientation relative to $\langle xyz \rangle$. Another quantity of interest is the total magnetization $\vec{M}[\vec{e}] = \sum_x \vec{e}_x$. The partition function of a $\theta$-vacuum is given by

$$Z(\theta) = \int \mathcal{D}\vec{e} \, \exp(-S[\vec{e}] + i\theta Q[\vec{e}]) = \prod_x \int_{S^2} d\vec{e}_x \, \exp(-S[\vec{e}] + i\theta Q[\vec{e}]). \quad (3)$$

We also consider the topological susceptibility $\chi_t(\theta) = (\langle Q^2 \rangle_\theta - \langle Q \rangle_\theta^2)/V$ and the magnetic susceptibility $\chi_m(\theta) = (\langle \vec{M}^2 \rangle_\theta - \langle \vec{M} \rangle_\theta^2)/V$. Using finite size scaling one can extract critical exponents from $\chi_t(\theta)$ and $\chi_m(\theta)$.

The Wolff cluster algorithm is a very powerful tool that has been used to beat critical slowing down in numerical simulations of $O(N)$-models [6]. Each step of the algorithm begins with the selection of a random direction characterized by a unit vector $\vec{r}$. Then the spin components in this direction are updated by forming spin clusters which can only be flipped collectively. When flipped, a spin $\vec{e}_x$ changes to $\vec{e}_x{'} = \vec{e}_x - 2(\vec{r} \cdot \vec{e}_x)\vec{r}$. To define the clusters, nearest neighbor spins $\vec{e}_x$ and $\vec{e}_y$ are connected by a bond with probability

$$p = 1 - \min[1, \exp(s(\vec{e}_x, \vec{e}_y) - s(\vec{e}_x{'}, \vec{e}_y))]. \quad (4)$$

With our definition of the action in eq.(1) a bond is always put if a spin flip would increase the relative angle of two neighboring spins beyond $2\pi/3$. Spins which are connected by bonds belong to one cluster. In the multi-cluster algorithm all spins of a cluster are flipped simultaneously with probability 1/2. The clusters are completely independent. This allows the construction of improved estimators for quantities like $\chi_m(0)$ [6].

Here we use the cluster algorithm to update the $\theta$-vacuum of the $O(3)$-model. Actually we update at $\theta = 0$ and include the extra Boltzmann factor $\exp(i\theta Q)$ in the measured observables. This requires a very precise determination of the topological charge distribution

$$p(Q) = \int \mathcal{D}\vec{e}\; \delta_{Q,Q[\vec{e}]}\; \exp(-S[\vec{e}]). \tag{5}$$

The partition function is then given by $Z(\theta) = \sum_Q p(Q)\exp(i\theta Q)$. Observables are computed separately in each topological sector

$$\langle \mathcal{O} \rangle_Q = p(Q)^{-1} \int \mathcal{D}\vec{e}\; \delta_{Q,Q[\vec{e}]}\; \mathcal{O}[\vec{e}] \exp(-S[\vec{e}]), \tag{6}$$

and their $\theta$-vacuum expectation value is given by

$$\langle \mathcal{O} \rangle_\theta = Z(\theta)^{-1} \sum_Q p(Q) \langle \mathcal{O} \rangle_Q \exp(i\theta Q). \tag{7}$$

The distribution $p(Q)$ varies over many orders of magnitude. In particular, large charges are suppressed exponentially. Still, at large values of $\theta$ the contributions from the large charges are important and need to be determined precisely. In this respect a reweighting technique [7] using a trial distribution $p_t(Q)$ has been useful. The trial distribution should be as close as possible to the real distribution $p(Q)$. In general this is difficult to achieve because one has not yet determined $p(Q)$. In our case the improved estimator described below provides a good trial distribution $p_t(Q)$. Then one works with an effective action $S_{\text{eff}}[\vec{e}] = S[\vec{e}] + \ln p_t(Q[\vec{e}])$ and determines $p(Q) = p_t(Q) \int \mathcal{D}\vec{e}\; \delta_{Q,Q[\vec{e}]}\; \exp(-S_{\text{eff}}[\vec{e}])$. This method can be combined with the multi-cluster algorithm. The clusters are grown as usual. Flipping a specific cluster changes the topological charge from $Q$ to $Q'$. Hence we define the cluster charge as $(Q - Q')/2$. Therefore each cluster carries a half-integer topological charge. Due to the triangular lattice and our choice of action — which constrains the relative angles of neighboring spins — the cluster charges are uniquely defined, i.e. the charge of a specific cluster is independent of the orientations of all other clusters. Moreover the cluster charges can be computed locally. These are highly nontrivial consequences of our construction. The trial distribution $p_t(Q)$ enters only when the clusters are flipped. The probability to flip a specific cluster is given by the Metropolis algorithm

$$p = \begin{cases} \min[1, p_t(Q)/p_t(Q')] & \text{if } p_t(Q') \neq p_t(Q), \\ 1/2 & \text{otherwise.} \end{cases} \tag{8}$$

Since the clusters are independent we can construct an improved estimator for the topological charge distribution. This is the essential ingredient of our method to simulate a $\theta$-vacuum. A configuration with $n$ clusters represents an ensemble of $2^n$ configurations related to the original one by cluster flips. The weight of a configuration depends on its topological charge and is given by $p_t(Q)^{-1}$. Knowing all cluster charges, it is a simple combinatoric exercise to construct the topological charge distribution for the ensemble of configurations analytically. This increases the statistics by a factor $2^n$. Typically there are $O(100)$ clusters in our configurations which results in a tremendous improvement. In particular, the partition function $Z(\pi)$ gets a positive contribution from each configuration. Hence there is no sign-problem unlike in other numerical methods. Similarly, we construct an improved estimator for the magnetic susceptibility. This is done separately for each topological sector. The orientation of the charged clusters determines the topological charge $Q$. When the neutral clusters are flipped we stay in the given topological sector. The improved estimator for $\langle \vec{M}^2 \rangle_Q$ includes all configurations that result from flipping the neutral clusters only, and it is otherwise identical with the standard improved estimator [6]. It turns out that most clusters are neutral and the improvement factor is still large.

We emphasize that the cluster formulation of the $O(3)$-model provides more than just an efficient numerical algorithm. The clusters represent physical objects that play a crucial role in the dynamics. Indeed we identify the clusters of charge $\pm 1/2$ with merons. This definition of merons is consistent with the semiclassical picture. In fact, an instanton contains two Wolff clusters, both with topological charge $1/2$. At $\theta = 0$ the clusters are independent and hence the merons form an ideal gas. The resulting disorder is responsible for the mass-gap of the theory. At $\theta = \pi$ the Boltzmann factor $\exp(i\theta Q)$ is $1$ for even $Q$ and $-1$ for odd $Q$. When a configuration contains a half-odd-integer charged cluster, flipping it changes $Q$ by an odd integer. The resulting configurations then have opposite Boltzmann weights and their contributions cancel in the partition function. Therefore, at $\theta = \pi$ only those configurations for which all clusters carry integer charges contribute to $Z(\theta)$. This means that the merons are now bound to pairs and can no longer disorder the system. Hence it is natural to expect that the mass-gap vanishes at $\theta = \pi$. This agrees with Haldane's conjecture and it provides an exact formulation of Affleck's dynamical picture.

We have performed numerical simulations with the meron-cluster algorithm at $g = \infty$ for volumes $V = 3L \times L = 18 \times 6$, $24 \times 8$, $30 \times 10$ and $36 \times 12$. For each lattice we have performed $10^7$ sweeps for measurements. The correlation length at $\theta = 0$ is about 2.8 lattice spacings. Fig.1 shows the topological charge distribution $p(Q)$ on the $36 \times 12$ lattice. Due to our reweighting technique we can easily generate a distribution that covers 25 orders of magnitude. Up to logarithmic corrections discussed below we find $\chi_t(\pi) \propto L$ and $\chi_m(\pi) \propto L$ which indicates a *second* order phase transition. For a first order transition the susceptibilities would grow proportional to $L^2$. In fact, at very strong coupling one expects a first order phase transition [11, 9]. Due to the constraint in our action of eq.(1) the strongest bare coupling we can consider ($g = \infty$) turns out to be too weak to reach this regime.

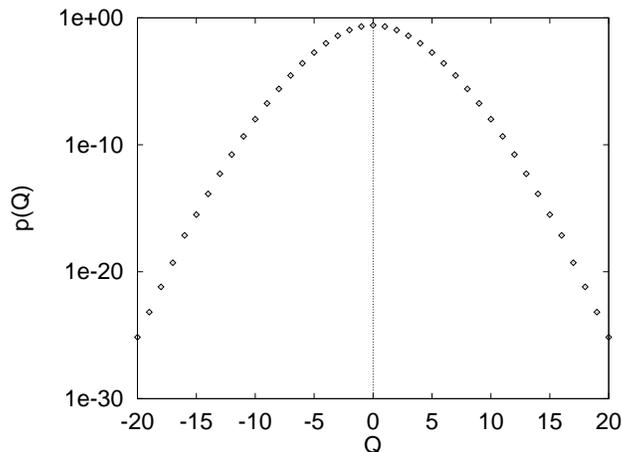

Figure 1: *The topological charge distribution $p(Q)$ on the $36 \times 12$ lattice.*

We have not run our algorithm at $g < 0$ although this may be feasible. It would be interesting if one could then reach the first order domain. Affleck et al. have conjectured that the critical theory at the second order phase transition is up to additional marginal operators a conformal field theory in the universality class of the $k = 1$ Wess-Zumino-Novikov-Witten model [3, 9]. Close to the phase transition the mass-gap of the infinite volume theory should behave as

$$m(\theta) = |\theta - \pi|^{2/3} |\ln(|\theta - \pi|)|^{-1/2}. \qquad (9)$$

We consider Fisher's finite size scaling variable $z = m(\theta)L$ which is a renormalization group invariant measure of the physical volume. Using the critical exponents of the Wess-Zumino-Novikov-Witten model with logarithmic corrections to scaling due to the additional marginal operators [3] one expects

$$\chi_t(\theta, L) = L \, (\ln L)^{-1/2} \, g_t(z), \quad \chi_m(\theta, L) = L \, (\ln L)^{1/2} \, g_m(z), \qquad (10)$$

close to the phase transition. Here $g_t(z)$ and $g_m(z)$ are *universal* functions. In figs.2, 3, $\chi_t(\theta, L)L^{-1}(\ln L)^{1/2}$ and $\chi_m(\theta, L)L^{-1}(\ln L)^{-1/2}$ are shown as functions of $z$. In both cases the data are close to a universal curve. The logarithmic corrections to scaling are important for this. Note that no fitting or free parameters are involved. This confirms the described scenario. Cluster diagnostics on the $36 \times 12$ lattice at $g = \infty$ shows that most clusters are neutral, 4 percent have charge $\pm 1/2$, 1 permille have charge $\pm 1$ and very few have larger charges. The average sizes of the 0, $\pm 1/2$ and $\pm 1$ charged clusters are 1.5, 10 and 32 lattice sites, respectively.

In conclusion we have identified a second order phase transition at $\theta = \pi$ with the critical exponents of the $k = 1$ Wess-Zumino-Novikov-Witten model. This confirms Haldane's conjecture in the framework of the $O(3)$-model. Identifying Wolff clusters of charge $\pm 1/2$ with merons we have provided a precise formulation of the meron picture. We have confirmed that the merons form integer charged pairs at the phase transition. The $O(3)$-model is identical with the $CP(1)$-model. All $CP(N)$-models have instantons and hence a $\theta$-vacuum. In the large $N$ limit the $CP(N)$-model has a first order phase transition at $\theta = \pi$ [11] where CP is spontaneously broken. One expects that this persists down to $N = 2$ [11, 9, 12] although a numerical study in the $CP(3)$-model (motivated by the strong CP-problem) seems to contradict this [13]. Unfortunately, cluster algorithms do not work well for $CP(N)$-models with $N \geq 2$ [14]. Lattice $CP(N)$-models with $N \geq 2$ are of special interest because their topological effects have a well defined continuum limit $g \to 0$. This is not the case for the $O(3)$-model [15]. Fortunately, in the condensed matter context of our investigation there is no need for scaling in the limit $g \to 0$. This limit corresponds to the classical limit $S \to \infty$ of the quantum spin system. The phase transition in $\theta$ is expected to persist for arbitrarily large spins, but there is no reason to expect scaling in $S$.

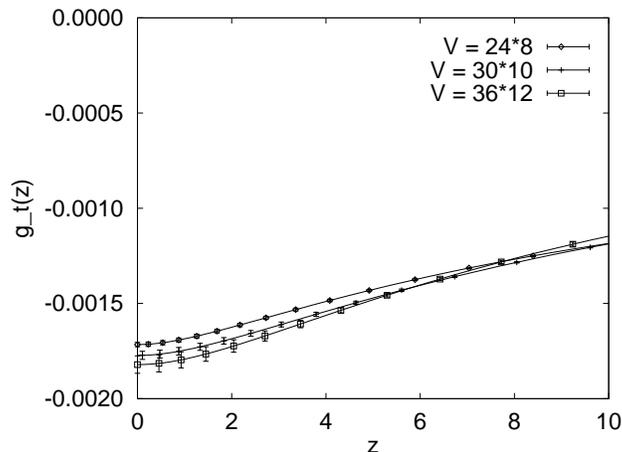

Figure 2: *Data for the universal function $g_t(z)$.*

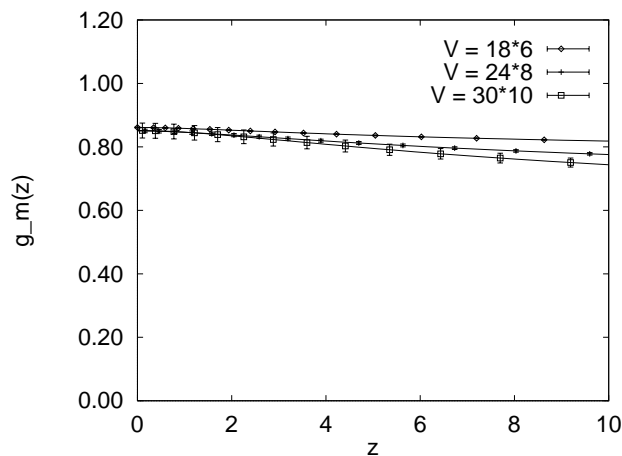

Figure 3: *Data for the universal function $g_m(z)$.*

Unfortunately the $\theta$-vacua in gauge theories cannot be simulated reliably at large $\theta$ because sufficiently powerful algorithms are not available. However, the combination of cluster and reweighting techniques is applicable to the Ising model. This is presently under investigation.

We are indebted to I. Affleck and M. Lüscher for very instructive communications.